\documentclass[12pt]{iopart}
\usepackage[dvips]{graphicx}
\usepackage{floatflt}

\newcommand{ \rts }{$\sqrt{s_{_{\rm NN}}}$}
\def \pp    {$p+p$} 
\def \auau  {Au+Au}
\def \dau   {$d$+Au}
\def \gev   {\ GeV/$c$}
\def \pt    {$p_T$}

\begin{document}

\title{Highlights from STAR}

\author{Kai Schweda for the STAR collaboration\footnote{see appendix for author list}}
\ead{koschweda@lbl.gov}

\address{Lawrence Berkeley National Laboratory, 
One Cyclotron Rd MS70R0319, Berkeley, CA 94720, USA}

\begin{abstract}
Selected results from the STAR collaboration are presented. 
We focus on recent results on jet-like 
correlations, nuclear modification factors of identified hadrons,
elliptic flow of multi-strange baryons $\Xi$ and $\Omega$, 
and resonance yields. First measurements of open charm production at RHIC
are presented.
\end{abstract}


\submitto{\JPG}


\section{Introduction}

The ultimate goal of ultra-relativistic nuclear collisions at RHIC 
is to identify and study matter with partonic -- quark and
gluon -- degrees of freedom. To address this important question, we study
\begin{itemize}
\item nuclear effects at intermediate and high transverse momentum \pt\ to probe
initial conditions and potentially partonic energy loss,
\item bulk properties of matter to probe the collisions dynamics, to detect
collective
motion among partons and finally to determine the equation of state of partonic 
matter.
\end{itemize}

STAR also has an active program on ultra-peripheral collisions, where the 
Au nuclei interact only via the long range electromagnetic force~\cite{UPC_rho}.
In collisions with polarized protons, we investigate the spin structure of
the nucleon~\cite{spin}.  

This paper is structured as follows. 
Section~2 discusses one example of ultra-peripheral collisions.
I will then exclusively focus on ultra-relativistic nuclear collisions.
In Sec.~3, recent results on jet-like 
correlations and nuclear modification factors at intermediate and high \pt\
are presented. Azimuthal anisotropy parameters are shown in Sec.~4. 
Section~5 discusses particle production and bulk properties. Section~6 
presents first measurements on open charm production at RHIC.

\section{Ultra-peripheral collisions}
\begin{figure}[tbh]
\begin{minipage}{0.49\textwidth}
\includegraphics[width=0.95\textwidth]{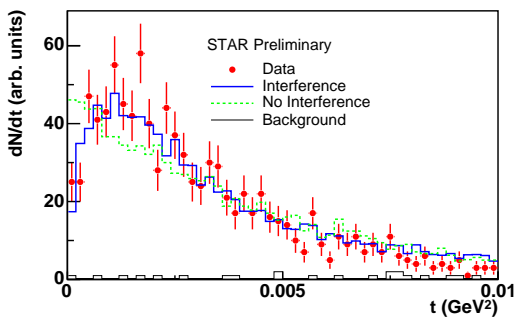}
\end{minipage}\hspace{-2.5cm}
\begin{minipage}{0.66\textwidth}\vspace{-8mm}
\caption{Raw spectrum for $\rho^0$ sample
at $0.1 < |y|<0.5$ from ultra-relativistic \auau\ collisions at \rts =200\ GeV. 
The points show our experimental
results. The solid histogram are results from calculations incorporating
coherent $\rho^0$ production, while the dashed histogram shows results from simulations 
without interference. The solid histogram with
very few counts displays the background contribution.}
\end{minipage}
\label{fig0}
\end{figure}

In ultra-peripheral collisions, the 
Au nuclei interact only via the long range electro-magnetic force.
Vector mesons are produced when a
photon from the electromagnetic field of one Au nucleus strikes 
the other one.  Production can occur at either nucleus; 
the two possibilities
interfere destructively, reducing production at small $p_T$
\cite{interf}.  \Fref{fig0} shows the raw $t_{\perp} = p_T^2$ spectrum
for $\rho^0$ photoproduction in $0.1 < |y|<0.5$ from
\auau\ collisions at \rts =200\ GeV.  STAR data
(points) are compared with calculations with and without the
interference.  The data drops at small $p_T$, showing that there is
interference between $\rho^0$ photoproduction at two well-separated
nuclei.
 
\vspace{-6mm}
\section{Measurements at intermediate and high $p_T$}
\begin{figure}[tbh]
\begin{minipage}{0.49\textwidth}
\includegraphics[width=0.95\textwidth,angle=-90]{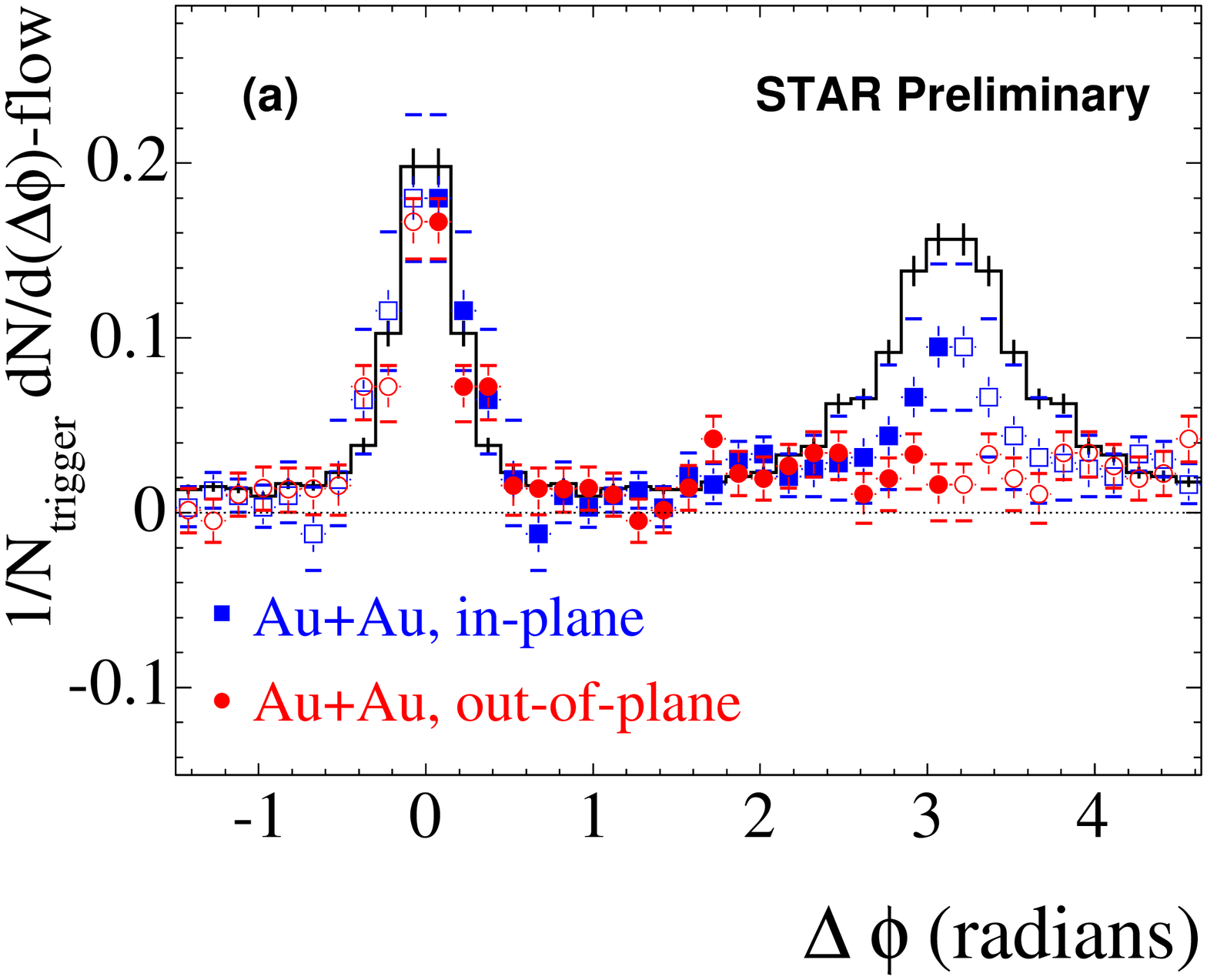}
\end{minipage}\hfill\hspace{2.5cm}
\begin{minipage}{0.49\textwidth}\vspace{-0.8cm}
\includegraphics[width=0.80\textwidth]{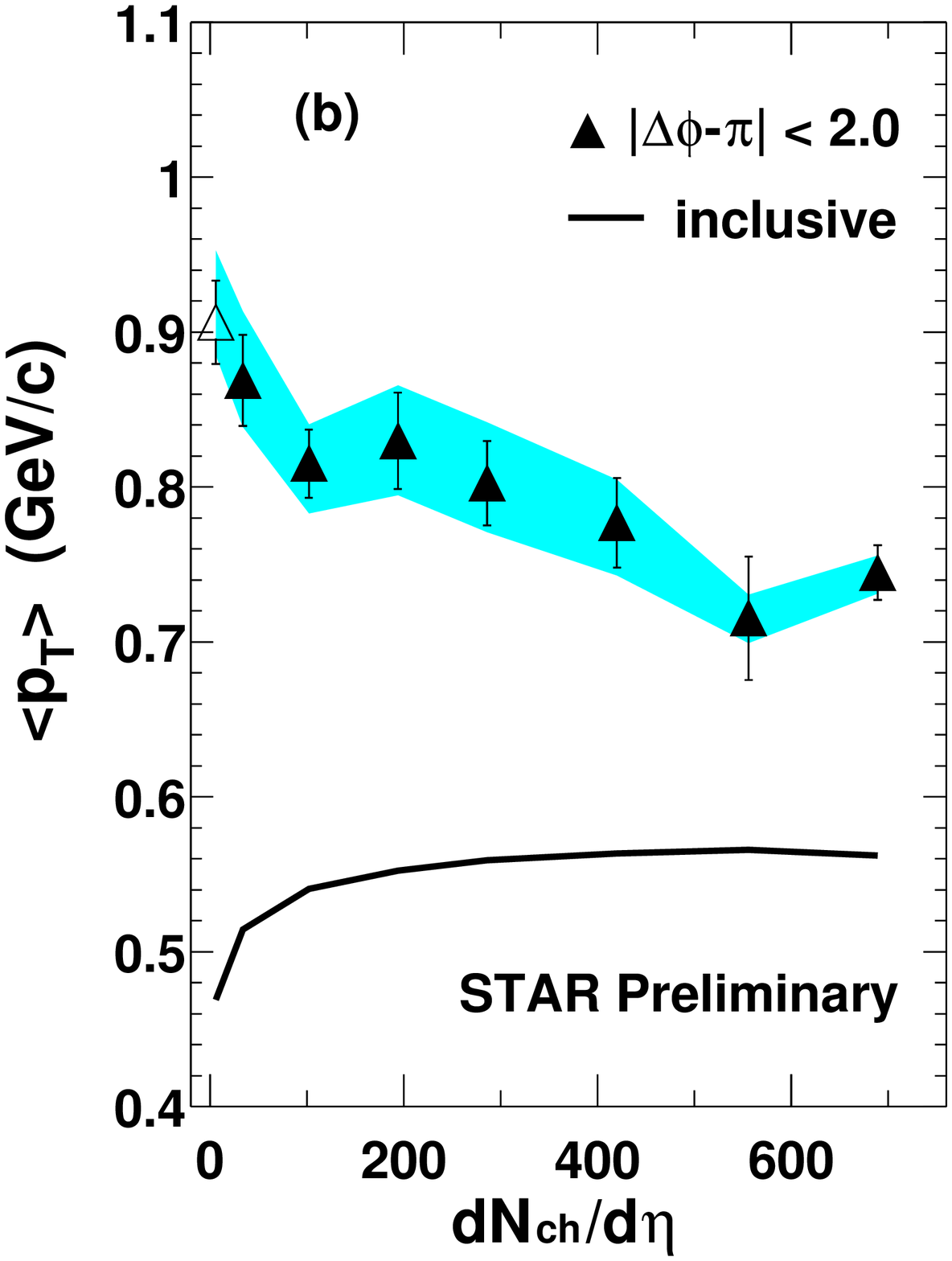}
\end{minipage}
\caption{(a) Azimuthal distribution 
of particles with respect to a trigger particle for \pp--collisions (solid line),
and mid-central \auau\ collisions within the reaction plane (squares) and
out-of-plane (circles) at \rts=200~GeV and (b) mean transverse momentum 
for particles around the away-side region as a function of number of 
charged particles. 
The solid line shows the mean transverse momentum of inclusive hadrons.}
\label{fig1}
\end{figure}

High transverse momentum hadrons provide an excellent probe 
of the high energy density matter created at RHIC~\cite{wang92}.
The measurement of particle yields and jet-like correlations at intermediate
and high $p_T$ have demonstrated that indeed a hot and dense medium is created
in \auau\ collisions at RHIC~\cite{hipt1, hipt2, hipt3, hipt4}. 
The solid line in \fref{fig1}(a) shows the azimuthal distribution 
of charged hadrons at 2\gev$\ <$\ \pt\ $<p_T^{\rm trig}$  
with respect to a trigger particle with 4\ $<$\ $p_T^{\rm trig}<6$\gev\ for 
\pp\ collisions at \rts=200~GeV. This distribution exhibits an 
enhancement around 
$\Delta \phi\sim 0$ which is typical of jet production, and around 
$\Delta \phi\sim\pi$ typical of di-jet events. 

On the other hand,
azimuthal distributions from mid-central \auau\ collisions for particles
within the reaction plane ($|\phi|<\pi/4$) and out-of the reaction plane 
($|\phi|>\pi/4$) are shown by squares and circles,
respectively~\cite{tang_QM}. 
Contributions from elliptic flow and pedestals have been subtracted.
Here, $\phi$ 
is the particle angle with respect to the reaction plane, while $\Delta \phi$ is
the particle angle with respect to the trigger particle. To increase statistics,
the results from \auau\ collisions are measured for $|\Delta\phi|$ 
(solid symbols) and were reflected about $|\Delta \phi| = 0$ 
and $|\Delta\phi|=\pi$ (open symbols).
All three distributions exhibit an enhancement in the near-side region. However,
the \auau\ results show a relative suppression in the away-side region 
which is stronger in the out-of-plane direction. The observed suppression
is consistent with the jet quenching scenario, where fast partons or their 
hadronic fragments lose energy due to interactions with the medium created 
at RHIC. The pathlength in the initial
overlap region of both Au nuclei is larger in the out-of-plane direction than
in-plane. Therefore, in the jet-quenching scenario one would expect a 
larger suppression in the out-of-plane direction which is consistent with 
our experimental data. 

\Fref{fig1}(b) shows the mean transverse momentum of charged hadrons with
$0.15<$\ \pt\ $<4.00$\gev\ associated with a trigger particle around the
away-side region~\cite{wang_QM} as a function of increasing centrality, as measured by the
number of charged particles. The background contribution  
was determined by an event--mixing technique and has been subtracted.
The mean transverse momentum is decreasing 
with centrality indicating a steady softening of the spectrum in the 
away-side region. In contrast,
the mean transverse momentum of inclusive hadrons is increasing due to the 
larger amount of transverse radial flow developed in more central collisions.
This is shown by the solid line in \fref{fig1}(b). 
The observed softening in the away-side region suggest that energy originally
carried by fast partons is largely dissipated in the collision medium.

Besides correlation with respect to a single trigger particle, STAR
has measured two-particle correlations locally on $\eta$--$\phi$ for charged
hadrons with \pt\ $<$\ 2.0\gev~\cite{appel_QM}. Even at these relatively 
small momenta,
we observe enhancements in the away-side and same-side region as well as a
suppression of the away-side to same-side amplitude ratio as a function
of centrality.

Models incorporating initial state parton saturation effects~\cite{dima_dau} 
also predict considerable suppression of hadron yields at high \pt. 
In our \dau\ control measurement, where a hot, dense medium is not
produced, we do not observe any suppression of yields in 
the \pt\ range 2--7\gev\ \cite{hipt5}. Therefore, our data strongly 
suggest that the observed suppression in \auau\ collisions is due to 
interactions in the medium. 

\begin{figure}[tbh]
\vspace{-0.3cm}
\centering\mbox{
\includegraphics[height=0.51\textwidth]{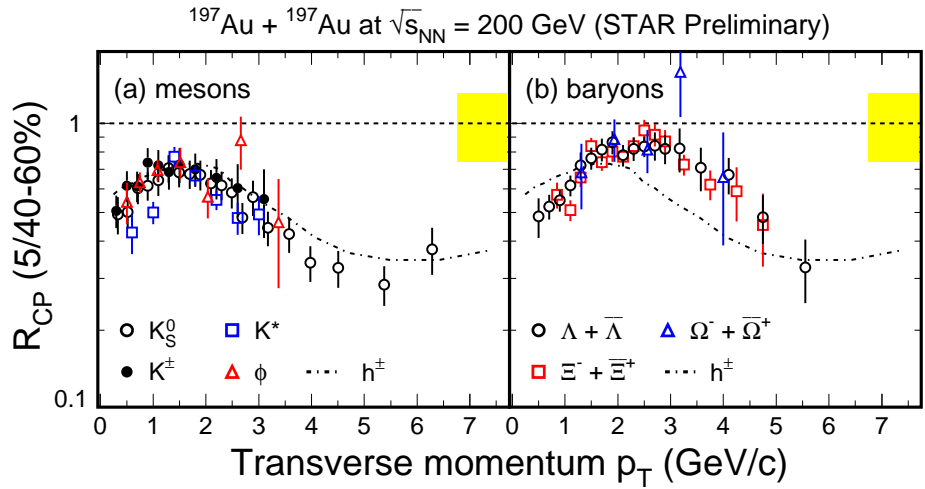}}
\vspace{-1.2cm}
\caption{(a) Nuclear modification factors of identified mesons and 
(b) identified baryons from \auau\ 
collisions at \rts=200~GeV. The dashed lines show experimental results for
charged hadrons. Peripheral \auau\ collisions were used as 
a reference.}
\label{fig3}
\end{figure}
Nuclear modification factors from 5\% most central \auau\ collisions 
at \rts=200~GeV~\cite{paul_04,lamont_QM, barnby_QM}  are shown in \fref{fig3} 
up to \pt\ $< 7$\gev\ for (a) identified mesons, namely 
K$^0_S$ (open circles), charged kaons (closed circles), 
K$^*$ (open squares), $\phi$ (open triangles) and (b) identified baryons 
$\Lambda$ and $\overline{\Lambda}$ (open circles), 
$\Xi$ and $\overline{\Xi}$ (open squares) and 
$\Omega$ and $\overline{\Omega}$ (open triangles).
The dashed lines show experimental results for
charged hadrons. 
As a reference, yields from
peripheral (40--60\%) \auau\ collisions were used. The hatched bands show Glauber 
model expectations for scaling of the yields with 
the number of binary collisions (\pt\ $>$\ 7.0\gev)
and their respective systematic uncertainties.  
In the intermediate \pt\ range at 2\ $<$\ \pt\ $<6$\gev,
there seem to be two groups: Mesons fall
into a common band while baryons 
show consistently larger values. Results for $\phi$ and 
$\Omega(\overline{\Omega})$ are still inconclusive due to low statistics.

Quark coalescence or recombination models~\cite{denes_03, fries_03,greco_03, fries_QM} for hadron formation offer an
elegant explanation for the experimentally observed 
dependence on the number of constituent quarks. Within some of 
these models~\cite{fries_03}, hadrons at 
intermediate \pt\ are dominantly formed by coalescing quarks stemming from
a thermalized parton system, i.e.\ a quark gluon plasma. 
Therefore, measurements at intermediate \pt\ might reveal information on  
partonic bulk matter.

\section{Azimuthal Anisotropy}
\begin{figure}[hbt]
\vspace{-0.4cm}
\centering\mbox{
\includegraphics[width=0.9\textwidth]{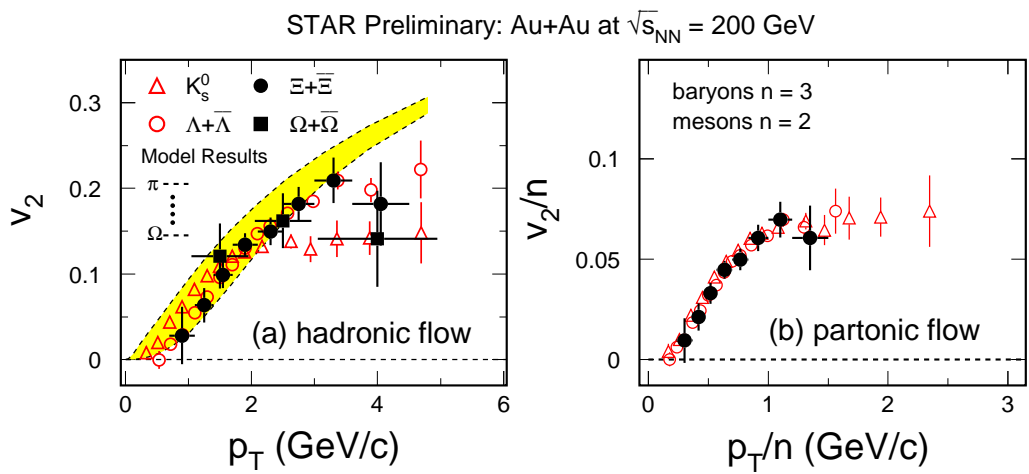}}
\vspace{-0.8cm}
\caption{(a) Measured azimuthal anisotropy parameters $v_2$ of strange hadrons
as a function of transverse momentum \pt\ for minimum bias \auau\ collisions at
\rts=200~GeV. 
The hatched band indicates typical results
from hydrodynamical calculations from the pion mass (upper dashed curve)
to the $\Omega$ mass (lower dashed curve) and (b) 
azimuthal anisotropy parameters $v_2/n$ versus \pt$/n$, with
$n$ the number of constituent quarks of the corresponding hadron.}
\label{fig4}
\end{figure}

Azimuthal anisotropy of particle production is sensitive to the early stage
of ultra-relativistic nuclear collisions~\cite{ollitrault92,sorge97}.
\Fref{fig4}(a) shows measured azimuthal anisotropy parameters $v_2$ 
of the strange hadrons K$^0_S$ (open triangles),\
$\Lambda$ and $\overline{\Lambda}$ (open circles), 
$\Xi$ and $\overline{\Xi}$ (closed circles) and 
$\Omega$ and $\overline{\Omega}$ (closed squares) as a function
of transverse momentum \pt\ for minimum bias \auau\ 
collisions at \rts=200~GeV~\cite{castillo_QM}.
Strange baryons exhibit a significant amount of elliptic flow. 

The hatched band indicates typical results
from hydrodynamical calculations~\cite{pasi01} from 
the pion mass (upper dashed curve)
to the $\Omega$ mass (lower dashed curve). The calculations describe the
experimental data well over a large range in \pt, but systematically
overshoot the data at \pt\ $> 3$\gev.  
Quark coalescence models predict a universal scaling of elliptic flow parameters
versus \pt\ with the number $n$ of constituent quarks. \Fref{fig4}(b)
shows our 
experimental data with $v_2$ and \pt\ scaled by $n$. 
Our results agree with the predicted scaling within statistical uncertainties. 
Higher statistics data is needed to further probe quark coalescence models.
Of special interest is a high statistics $v_2$ measurement of $\phi$ and 
$\Omega$ which will allow for the unambiguous distinction between parton 
recombination and statistical hadro-chemistry to be the dominant process
in hadronization at intermediate \pt~\cite{nonaka_04a}. 
Note, that kaon fusion is unlikely 
to be the dominant production mechanism of $\phi$ mesons~\cite{geno_03}. 
Therefore, the $\phi$ meson might carry dominantly pre-hadronic information. 
In addition,   
elliptic flow measurements of the resonances $\rho$ and $K^*$ will offer
to disentangle the contribution of partonic and hadronic 
interactions~\cite{nonaka_04b}.

STAR also measured directed flow, $v_1$~\cite{tang_QM} and the higher
harmonics $v_4$ and $v_6$~\cite{v4v6, posk_QM}. 
For the first time at RHIC, we determined the sign of $v_2$ to
be positive, i.e.\ in-plane elliptic flow~\cite{tang_QM}. 

In addition, we note that 
two-pion Hanbury Brown--Twiss interferometry relative to the
reaction plane indicates that the system at pion kinetic freeze-out is still
elongated perpendicular to the reaction plane~\cite{STAR_HBT}.

\section{Bulk properties}

\begin{figure}[hb]
\begin{minipage}{0.4\textwidth}
\includegraphics[width=0.9\textwidth, angle=-90]
{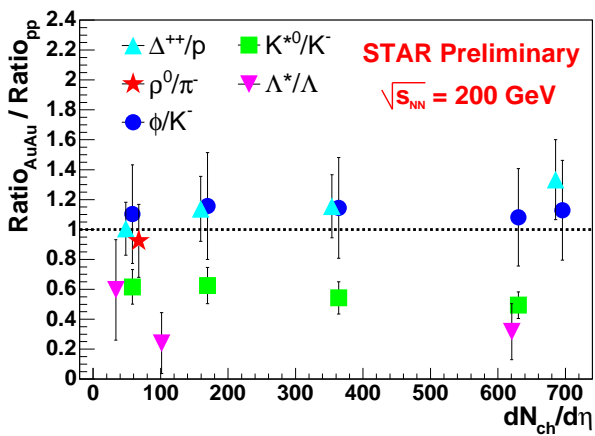}
\end{minipage}\hfill
\begin{minipage}{0.6\textwidth}\vspace{-2.7cm}
\caption{Ratios of resonances compared to their corresponding stable 
particle with identical constituent quark content for \auau\ collisions at
\rts=200~GeV as a function of centrality. All ratios have been normalized to 
unity for \pp\ collisions, as indicated by the dashed line.
}
\end{minipage}
\label{fig6}
\end{figure}

Experimental ratios of stable particles have been successfully described within
the statistical model using a temperature parameter $T_{ch}=177\pm7$~MeV and 
a baryo-chemical potential $\mu_b=29\pm6$~MeV in most central \auau\ collisions at
\rts=200~GeV~\cite{pbm, olga_04}. However, the kinetic freeze-out 
temperature has been found
to be much lower, $T_{kin}=89\pm10$~MeV~\cite{olga_04}.  
Resonances
and their hadronic decay daughters might participate in (quasi-)elastic 
interactions at the late hadronic stage and therefore probe the evolution of
the medium until kinetic freeze-out.
\Fref{fig6} shows ratios of $\Delta^{++}, \rho^0, \phi$, K$^*$ and 
$\Lambda$(1520) compared to their corresponding stable particle
with identical constituent quark content for \auau\ collisions at
\rts=200~GeV as a function of centrality~\cite{markert_QM}. 
All ratios have been normalized to 
unity for \pp\ collisions, as indicated by the dashed line.
Clearly, the ratios of K$^*$/K$^-$ and 
$\Lambda(1520)/\Lambda$ are modified in \auau\ collisions. This indicates that
there is an active hadronic stage after chemical freeze-out.

\section{Heavy Flavor Measurements in \pp\ and $d+$Au Collisions}
\begin{figure}[thb]
\centering\mbox{
\hspace{-0.6cm}
\includegraphics[width=0.5\textwidth]{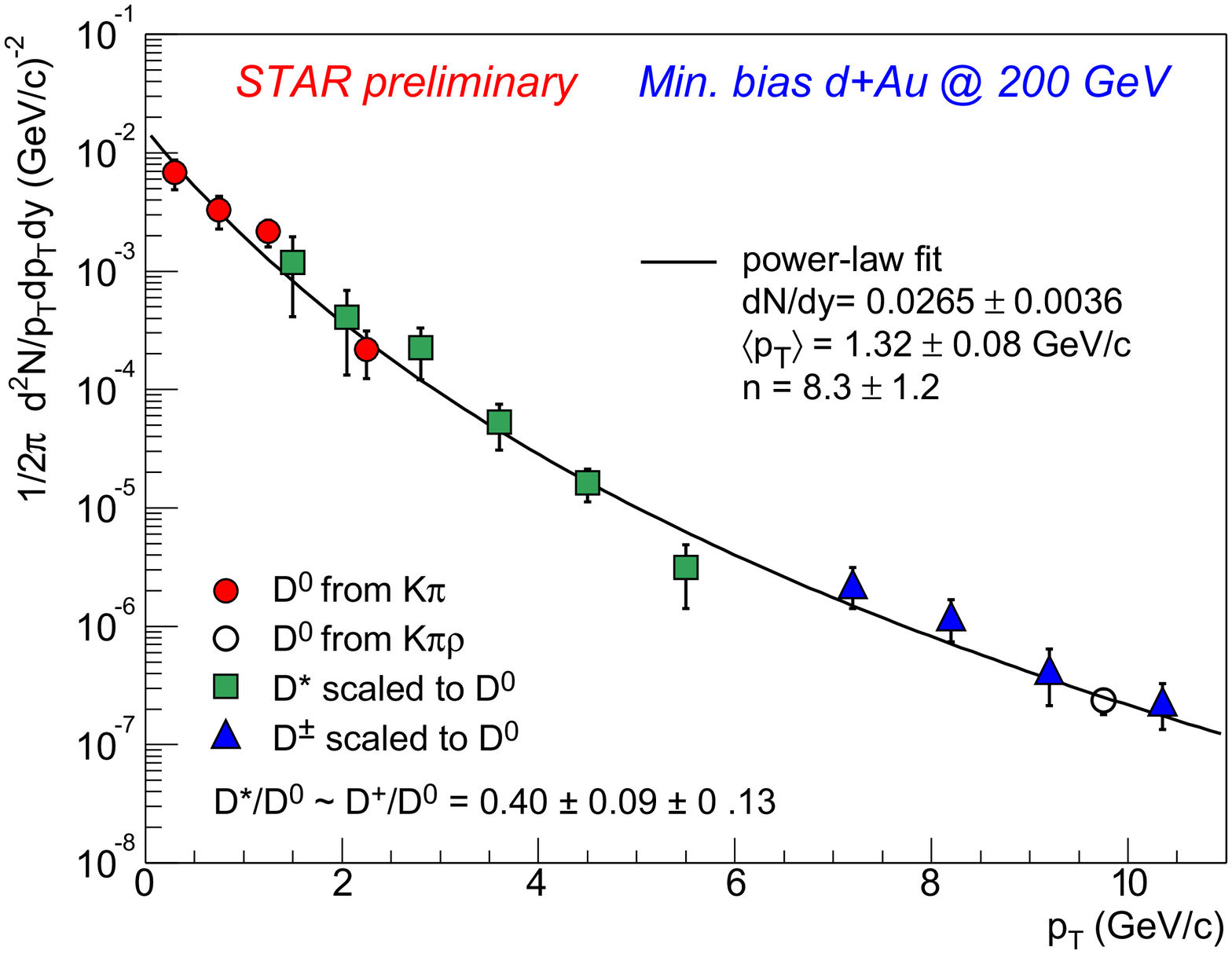}
\includegraphics[width=0.525\textwidth]{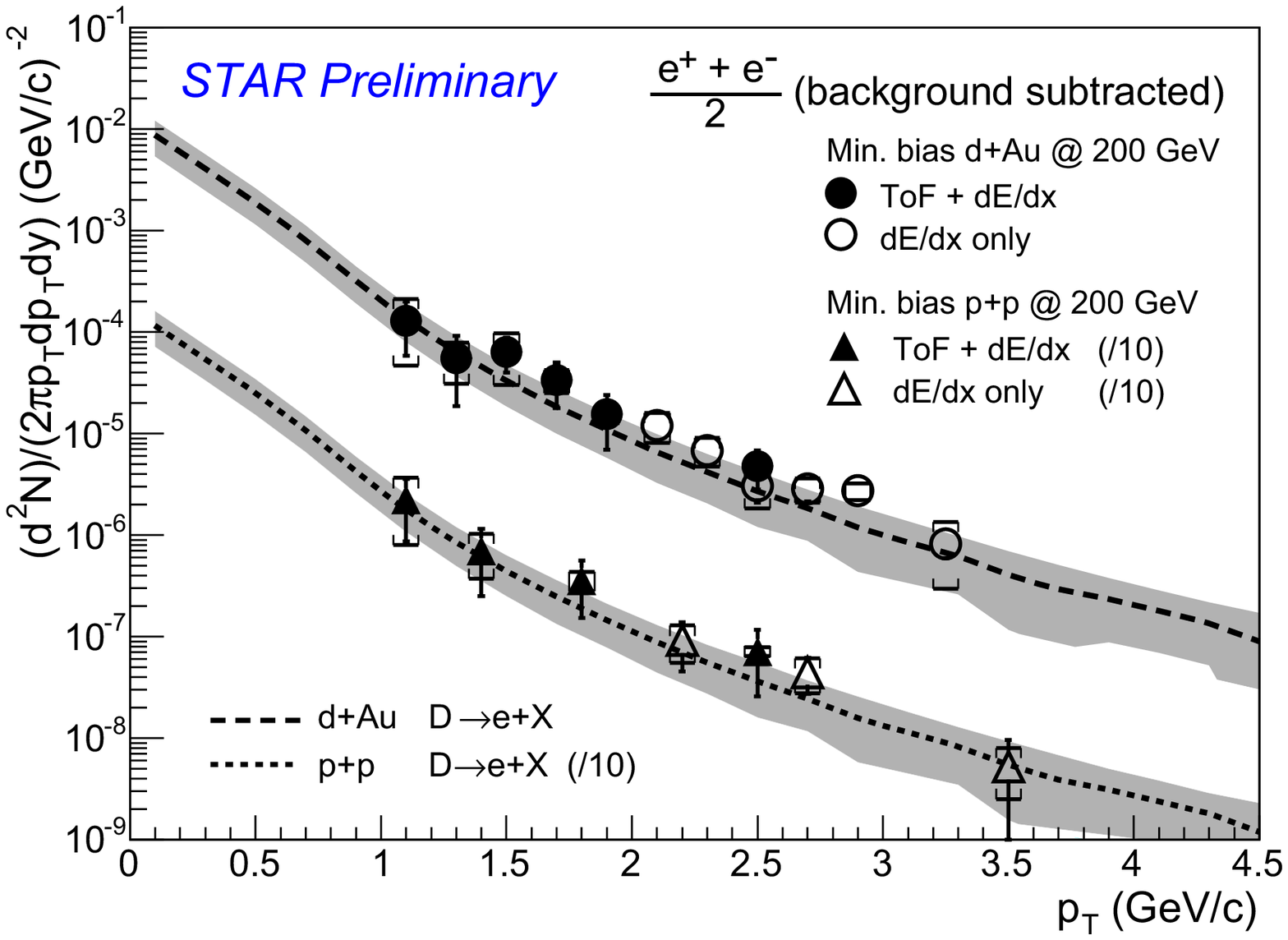}}
\caption{Left: invariant yield
of D$^0$~(circles), D$^*$~(squares) and D$^\pm$~(triangles)
from \dau\ collisions at \rts=200~GeV as a function
of transverse momentum. Right: invariant yield of summed electrons and positrons 
as a function of transverse momentum from \dau\ (upper points) and 
\pp\ collisions (lower points). 
Background from electromagnetic decays has been subtracted.
The shaded bands indicate the expected summed electron and positron yield from our
D-meson measurements.}
\label{fig8}
\end{figure}

STAR has studied D-meson production at RHIC. D-mesons were identified by
calculating the invariant mass of hadronic decay daughter
candidates~\cite{tai_QM} and via electrons(positrons) from semi-leptonic
decays~\cite{ruan_QM, suaide_QM}. 
The left hand side of \fref{fig8} shows the invariant yield
of D$^0$~(circles), D$^*$~(squares) and D$^\pm$~(triangles)
from \dau\ collisions at \rts=200~GeV
as a function of transverse momentum \pt\ in the range from 0.2 to 11.0\gev.
The yields of D$^*$ and D$^\pm$ have been arbitrarily 
scaled assuming identical spectral shapes of all D-mesons measured. 
Our experimental yields correspond to a total $c\overline{c}$ production 
cross section per nucleon--nucleon collision
of $1.18\pm0.21{\rm\ (stat.)}\pm0.39{\rm\ (syst.)}$~mb. Results from next-to-leading
order perturbative QCD calculations~\cite{ccbar} predict significantly smaller numbers.

Production of heavy flavor, i.e.\ charm and beauty was measured
in the electron(positron) decay channel. The right hand side of \fref{fig8}
shows the invariant yield of summed electrons and positrons~\cite{ruan_QM} as a function of transverse
momentum \pt\ from $d+Au$ (upper points) and \pp\ collisions (lower points). 
Background from electromagnetic decays has been subtracted.
For electron(positron) identification, information on the specific 
energy loss in the TPC-gas (dE/dx) and time of flight (ToF) was used.
The shaded bands indicate the expected electron and positron yield from our
D-meson measurement. Within systematic errors
of $\approx$30\%, both measurements are in good agreement.
The Barrel Electromagnetic Calorimeter (BEMC) allowed for another independent
method of electron(positron) identification~\cite{suaide_QM} 
up to \pt\ $<7$\ \gev. Decays of B-mesons are expected to dominate 
the electron(positron) spectrum at \pt\ $>$\ 5\gev~\cite{suaide_QM}.    
All three measurements, namely 
direct identification of D-mesons by their invariant mass, identifying 
decay electrons by dE/dx and/or ToF and the BEMC lead to consistent results.

\section{Summary}
In summary, the suppression of hadron production at intermediate and 
high \pt\ in central \auau\ collisions and the \dau\ control measurement
demonstrate that there are interactions in the medium at the early, most
likely partonic stage. As a consequence, bulk matter created at RHIC exhibits
strong collective expansion, e.g. large values of elliptic flow.
The substantial amount of elliptic flow observed for 
multi-strange hadrons and the successful description within quark coalescence
models suggest that collectivity is indeed built up at the partonic level. 
We presented first measurements on open charm production at RHIC. 

In the future, we need measurements on nuclear modification factors and 
jet-like correlation of particles carrying heavy flavor (c-- and b--) quarks to
establish microscopic probes of different mass in order to further 
characterize the medium created at RHIC. 
A high statistics elliptic flow measurement of all hadrons, especially 
$\phi$, $\Omega$ and the resonances $\rho$ and K$^*$ will enable us
to  quantify parameters of partonic collectivity and 
disentangle partonic from hadronic contributions. 
The ongoing high statistics 
run at RHIC and our increasingly extending time of flight capabilities 
as well as an inner $\mu$Vertex detector for heavy-flavor identification
which is presently under development
will help us in achieving these goals. 

\vspace{1cm}

\end{document}